\documentclass[aps,prc,reprint,showpacs,groupedaddress,onecolumn]{revtex4-1}
\usepackage{graphicx,color}
\usepackage{dcolumn}
\usepackage{amssymb}
\usepackage{amsfonts}
\usepackage{color}
\usepackage[usenames,dvipsnames,svgnames,table]{xcolor}

\usepackage{epsfig}
\usepackage{bm}

\newcommand{\nc}{\newcommand}

\def\beq{\begin{equation}}
\def\eeq{\end{equation}}
\def\beqa{\begin{eqnarray}}
\def\eeqa{\end{eqnarray}}

\nc{\h}{\mathbf{h}}
\nc{\dg}{\mathbf{d}}
\def\p{{\sf p}}

\def\k{{\sf k}}

\nc{\bk}{{\bf k}}
\nc{\bp}{{\bf p}}
\nc{\bpp}{{\bf p}'}

\nc{\bq}{{\bf k}}
\nc{\bpjk}{{\bf p}_{jk}}
\nc{\bqi}{{\bf k}_i}
\nc{\bpki}{{\bf p}_{ki}}
\nc{\bqj}{{\bf k}_j}
\nc{\bpij}{{\bf p}_{ij}}
\nc{\bqk}{{\bf k}_k}

\nc{\bki}{{\bf k}_i}
\nc{\bkj}{{\bf k}_j}
\nc{\bkk}{{\bf k}_k}
\nc{\pjk}{p_{jk}}
\nc{\qi}{k_i}
\nc{\pki}{p_{ki}}
\nc{\qj}{k_j}
\nc{\pij}{p_{ij}}
\nc{\qk}{k_k}
\nc{\ki}{k_i}
\nc{\kj}{k_j}
\nc{\kk}{k_k}
\nc{\bera}{\langle}
\nc{\ket}{\rangle}
\nc{\bpinr}{\boldsymbol \pi_{nr}}
\nc{\btpinr}{\tilde{\boldsymbol \pi}_{nrx}}

\nc{\bpi}{\boldsymbol \pi}
\nc{\btpi}{\tilde{\boldsymbol \pi}}

\nc{\pinr}{ \pi_{nr}}
\nc{\tpinr}{\tilde{\pi}_{nr}}
\nc{\bQ}{{\bf Q}}
\nc{\bv}{{\bf v}}

\nc{\tpi}{\tilde{\pi}}
\nc{\om}{\omega_m}

\nc {\IR} [1]{\textcolor{red}{#1}}
 \nc {\IB} [1]{\textcolor{blue}{#1}}
 \nc {\IP} [1]{\textcolor{magenta}{#1}}

{

\begin{document}

\title{The Relativistic Three-Body Bound State in a 3D Formulation}

\author{M. R. Hadizadeh, Ch. Elster}
\affiliation{Institute of Nuclear and Particle Physics and Department of Physics
and Astronomy, Ohio University,
Athens, OH 45701, USA}

\author{W. N.\ Polyzou}
\affiliation{Department of Physics and Astronomy,
The University of Iowa, Iowa City, IA 52242, USA}

\date{\today}

\begin{abstract}
\noindent
{\bf Background:} The relativistic three-body problem has a long tradition
in few-nucleon physics. Calculations of the triton binding energy based on
the solution of the relativistic Faddeev equation in general lead to a weaker
binding than the corresponding non-relativistic calculation. \\
{\bf Purpose:} In this work we solve 
for the three-body binding energy as well as the wave function and its momentum
distribution. The effect of  
the different relativistic ingredients are studied in detail.
\\
{\bf Method:} Relativistic
invariance is incorporated within the framework of Poincar{\'e}
invariant quantum mechanics. The relativistic momentum-space Faddeev
equation is formulated and directly solved in terms of momentum vectors
without employing a partial-wave decomposition. \\
{\bf Results:} The relativistic calculation gives a three-body binding
energy which is about 3\% smaller than its non-relativistic counterpart. 
In the wave function, relativistic effects are manifested in the Fermi
motion of the spectator particle. \\
{\bf Conclusions:} Our calculations show that though the overall
relativistic effects in the three-body bound state are small, 
individual effects by themselves are not necessarily small and must be 
taken into account consistently.
\end{abstract}


\maketitle


\setcounter{page}{1}

\section{Introduction}
\label{intro}

We solve the relativistic three-nucleon bound-state problem and
compare the resulting wave functions to the corresponding
non-relativistic bound-state wave functions. While the wave functions
themselves are not observable, the difference between the relativistic
and non-relativistic wave functions provide useful information about
which observables might be sensitive to the difference.
Before going into details we need to define what we consider
the relativistic
three-nucleon problem, what we mean by relativistic effects,
and  summarize what has been learned from previous work on this
problem. 

In discussing the three-nucleon problem we limit our considerations to
an idealized system modeled on a three-nucleon Hilbert space. This
limitation allows us to make meaningful comparisons with the
non-relativistic problem. The difference between a relativistic and
non-relativistic model is the underlying symmetry group of the theory.
For relativistic models the symmetry group is the Poincar\'e group and
for non-relativistic models it is the Galilean group. Symmetries of a
quantum theory preserve observables, (i.e. probabilities, expectation
values and ensemble averages). This ensures the invariance of these
observables in all inertial reference frames. In the relativistic
case the inertial frames are related by Poincar\'e transformations
while in the non-relativistic case inertial frames are related by
Galilean transformations. Symmetries in a quantum theory are
implemented by unitary or anti-unitary transformations. In the
relativistic case the dynamics is implemented by a unitary projective
representation of the Poincar\'e group \cite{Wigner:1939cj}. In the
non-relativistic case the dynamics is given by unitary projective
representation of the central extension \cite{Bargmann:1954gh} of the
Galilei group. Neither of these symmetries impose strong constraints
on the dynamics. Normally the dynamics is formulated in a particular
frame, e.g. the laboratory frame, the center-of-momentum (c.m.) frame,
etc. The symmetry only ensures that the results are consistent in
frames related to this particular frame by the symmetry transformations.

A second related constraint is cluster separability. In the
three-body system cluster separability means that isolated one- and
two-body subsystems should exhibit the same symmetries as the system itself.
This is needed to ensure that either special or Galilean relativity can
be tested on isolated subsystems. To understand the role of this
condition assume, for example, that the two- and three-body dynamics
are formulated in the two- and three-body rest frames respectively.
The two-body subsystems in the three-body rest frame are not generally
in the two-body rest frame. However, if the model satisfies cluster
properties, then the two-body symmetry transformation determines how to
transform the two-body subsystem from its rest frame to the
three-nucleon rest frame. This embedding is different for the
Galilean and Poincar\'e symmetry groups, and is the
source of the relativistic effects that will be studied in this work.

One feature of realistic nucleon-nucleon (NN) interactions is that
while they are formally motivated by e.g. meson-exchange models, when
cast in a non-relativistic two-nucleon Hamiltonian, the model is
adjusted so that the predicted NN observables agree with the
experiment with a $\chi^2$ per degree of freedom close to
1~\cite{Wiringa:1994wb,Machleidt:2000ge}. The experimental data is
consistent with special relativity while the non-relativistic
calculation is consistent with a Galilean symmetry. There is an
immediate question about what is being compared to the data to obtain
the quoted $\chi^2$. The answer depends on how the analysis is
performed. Normally laboratory frame data is correctly transformed to
the c.m. frame using a Lorentz transformation. The correctly
transformed scattering data is then compared to the non-relativistic
c.m. scattering solutions. Parameters of the interaction are fine
tuned to achieve agreement with the data. In this case phase shifts
are identified as functions of the relativistic and non-relativistic
relative momenta. This means that in any other frame the experimental
and computed cross sections will no longer be identical functions of
laboratory energy.

The important consequence of this is that in the preferred
c.m. frame the non-relativistic two-body calculation gives
the experimental result. It is also possible to introduce
relativistic NN interactions that are consistent with the
same scattering data. If one were to take the non-relativistic limit
of the relativistic model the scattering observables would change as a
result of the approximation and would not agree with the results of a
non-relativistic model that is fit to the same experimental data. 
This aspect of realistic NN interaction must be taken into account when 
interpreting relativistic corrections in the three-body problem.

In this work we construct the relativistic NN interaction so that the
rest-frame relativistic and non-relativistic NN wave functions and
phase shifts are identical. Cluster properties determine how these
two-body models should be embedded in the three-nucleon Hilbert space.
Relativistic effects are due entirely to the different ways that these
two-body interactions appear in the three-nucleon problem in order to
satisfy cluster properties.

The resulting formulation of the relativistic three-nucleon problem
has the property that in the limit that the momenta are all small
compared to the nucleon masses, the relativistic Faddeev equation
reduces to the non-relativistic one, which justifies our
interpretation of the difference being attributed to ``relativistic
effects''. It is appropriate to think of the relativistic effects
being due to the difference between relativistic and non-relativistic
treatments of Fermi motion, which involve subsystem Galilean or
Poincar\'e boosts respectively.

In this initial work the non-relativistic NN potential is a
Malfliet-Tjon V~\cite{Malfliet:1968tj} type interaction. The formal
definition of the phase and wave-function equivalent relativistic NN
interactions that we use was given by Coester, Pieper and
Serduke~\cite{Coester:1975zz}. The construction of the corresponding
relativistic NN transition operators was given
in~\cite{Keister:2005eq} and successfully implemented
in~\cite{Lin:2007ck,Lin:2008sy,Lin:2007kg}. The two-body unitary
representation of the Poincar\'e group is formulated using a
construction given by B. Bakmjian and
L. H. Thomas~\cite{Bakamjian:1953kh}. The corresponding three-body
unitary representation of the Poincar\'e group that satisfies
$S$-matrix cluster properties was introduced by
Coester~\cite{Coester:1965zz}. Note that while it is possible to
realize cluster properties of the unitary representation of the
Poincar\'e group~\cite{Sokolov:1977,Coester:1982vt,Keister:1991sb},
for the three-body problem it is sufficient and far simpler to formulate
a model where the $S$-matrix clusters properly. The difference is that
cluster properties of the unitary representation of the Poincar\'e
group requires three-body forces that are generated by the two-body
forces. In the formulation where only the $S$ matrix clusters the
required three body-forces are replaced by two-body interactions that
depend on the spectator momentum. The important relation is that both
formulations of the relativistic three-body problem give identical
three-body scattering observables.  They are related by an $S$-matrix 
preserving  unitary transformation that becomes the identity in 
the three-body rest frame \cite{Coester:1982vt}.

While many of the methods mentioned in the previous paragraph are
formulated using one of Dirac's forms of dynamics~\cite{Dirac:1949cp}, in our
calculations the form of dynamics is only relevant if we choose to
transform our results from the three-body rest frame to an arbitrary
frame. This can be done consistently in any form of dynamics
\cite{Polyzou:2010} .

In this work we solve the relativistic Faddeev equation using direct
integration~\cite{Elster:1998qv,Fachruddin:2000wv,Liu:2004tv}. This
method gives the three-body wave function directly, so it does not
have to be reconstructed using partial-wave methods. Part of our
motivation for using direct integration in the bound-state problem is
to provide test problems that can be compared to previous relativistic
partial-wave calculations. Ultimately these
calculations need to be extended to treat spin-dependent interactions 
and scattering at
relativistic energies, where direct-integration methods are
essential.

The first relativistic three-body calculations using the formalism
that we use in this paper were performed by Gl\"ockle,  Coester, and
Lee~\cite{Glockle:1986zz}. Their calculations used the same
Malfliet-Tjon V (MT-V interaction. However, they approximated the relativistic
interaction so the phase equivalence with the non-relativistic
interaction was only approximate and they used a partial-wave
expansion that was truncated to $s$-waves. They found a small
decrease in the binding energy due to relativistic effects.
Relativistic bound-state calculations with a realistic interaction
were performed by Kamada et al.~\cite{Kamada:2008xc}. These were fully
converged partial-wave calculations. The resulting binding energy
corrections were comparable to the corrections obtained by Gl\"ockle, 
Coester and Lee. Three-body scattering calculations using a realistic
interaction and including a three-nucleon force were also performed
using partial-wave methods for energies up to 250 MeV
\cite{Witala:2009zzb}. These calculations exhibited relativistic effects in the
breakup observables and large-angle elastic scattering showed evidence
of missing degrees of freedom. Direct integration scattering
calculations were successfully performed using the MT- V
interaction for energies up to 2~GeV
\cite{Lin:2007ck,Lin:2008sy,Lin:2007kg}. These also exhibited strong
relativistic effects in certain breakup observables and demonstrated
the value of direct integration methods at higher energies. Our long
term goal is to perform few-Gev scale scattering calculations with
realistic interactions. This requires using direct integration with
realistic interactions. Since there are no such calculations the
first step it to establish that the methods works for the bound-state
problem, where the results can be compared to partial-wave
calculations. This work is a precursor to including a relativistic
treatment with realistic interactions using direct integration methods. 

This work addresses two omissions of the body of work discussed above.
First there are no relativistic bound state calculations that have
utilized direct integration methods. Second, while the three-body
binding energy has been computed, there is no published work comparing
the relativistic and non-relativistic bound-state wave functions.
While wave functions are not directly observable, observables are
sensitive to the structure of the wave functions and differences in
the relativistic and non-relativistic wave functions are responsible
for relativistic effects in three-body observables.


\section{The Poincar\'e invariant Faddeev equation}
\label{sec2}
\subsection{Kinematic Variables}

In both the relativistic and non-relativistic three-body problem
Poincar\'e or Galilean symmetry relates the state of the system in its
rest frame to its state in a general frame. The dynamics is usually
formulated in the rest frame of the system. Thus it is
useful to formulate the three-body problem using variables that
describe the momenta of particles in the three-body rest frame. 
The two-body subsystem, on the other hand,  uses variables that describe the
momenta of particles in the rest frame of the subsystem. The relevant
variables are defined by boosting the single-particle momenta to the
three-body rest frame, and then boosting two-body subsystem
momenta to the subsystem rest frame. Using Galilean boosts results in the
Jacobi momenta that are used in non-relativistic three-body
calculations~\cite{Gloecklebook}. Replacing the Galilean boosts by
Lorentz boosts leads to relativistic Jacobi momenta. In both cases
these are changes of variables from single-particle momenta 
to variables that are more convenient in three-body applications.

In what follows we assume that all nucleons have the same mass, $m$. We
denote the single-particle four momenta by $p_i^\mu$.  We define total
four momentum of the non-interacting three-body system $P^{\mu}:=
\sum_{i=1}^2 p_i^{\mu}$ and its invariant mass $M_0^2 = - P^{\mu}
P_{\mu}$.  Relativistic Jacobi momenta are constructed by 
first boosting the ${p}_i^{\mu}$ to the three-body rest frame 
with a rotationless boost  
$\Lambda^{-1}  (\mathbf{P}/M_0)^{\mu}{}_{\nu}$,
   
\beq
k_i^{\mu} := \Lambda^{-1}  (\mathbf{P}/M_0)^{\mu}{}_{\nu} p_i^{\nu}.
\eeq
The vector components of  $k_i^{\mu}$ are
\beq
\bqi = \mathbf{p}_i  + {\mathbf{P}\over M_0} \left({\mathbf{P}
\cdot \mathbf{p}_i
\over M_0+ \sqrt{M_0^2 + \mathbf{P}^2}} - \omega_m ({p}_i)\right),
\label{2.7}
\eeq
where $\mathbf{P}$ is the three-vector part of $P^{\mu}$ 
and
$\omega_m({p}_i) = p_i^0 = \sqrt{m^2 + p_i^2}$ is the energy of the 
$i$-th particle.

The $\mathbf{k}_i$ are not independent. They satisfy 
\beq
\sum_{i=1}^3 \mathbf{k}_i =\mathbf{0}
\qquad
M_0 := \sum_{i=1}^3 \omega_m({k}_i) .
\label{2.7a}
\eeq

The other relativistic Jacobi momentum variables are obtained by 
boosting $k_i^{\mu}$ to the rest frame of the $ij$ pair.  
Following Eq.~(\ref{2.7}) we denote the four momentum
of the pair ($ij$) in the three-body rest frame by
$k^{\mu}_{ij}=
k^{\mu}_{i}+ k^{\mu}_{j}$, and the two-body invariant mass of the ($ij$)
subsystem 
by $m^2_{0ij} = - k^{\mu}_{ij} k_{\mu ij}$.  
The other relativistic Jacobi momenta are
defined by 
\beq
p_{ij}^{\mu} := \Lambda^{-1}  (\mathbf{k}_{ij}/m_{0ij})^{\mu}{}_{\nu} k_i^{\nu}.
\eeq
The vector components of $p_{ij}^{\mu}$ are 
\beq
\bpij  =  \bqi  
+  
{\mathbf{k}_{ij}\over m_{0ij}} \left({\mathbf{k}_{ij}\cdot \bqi
\over m_{0ij}+ \sqrt{m_{0ij} + k_{ij}^2}} - 
(\omega_m (k_i) )\right).
\label{2.11}
\eeq
The inverse of Eq.~(\ref{2.11}) is given as
\beq
\bqi =  \bpij   + 
{\mathbf{k}_{ij}\over m_{0ij}} + \left({\mathbf{k}_{ij}\cdot 
\mathbf{p}_{ij}
\over m_{0ij} \sqrt{m_{0ij} + k_{ij}^2}} + 
(\omega_m (p_{ij}) )\right).
\label{2.13}
\eeq
The pairs $(\mathbf{k}_k,\mathbf{p}_{ij})$ are the relativistic 
analogs of the usual Jacobi momenta.  If the Lorentz boosts 
$\Lambda^{-1}  (\cdot)^{\mu}{}_{\nu}$ are replaced by 
Galilean boosts these become the non-relativistic 
Jacobi momenta~\cite{Gloecklebook}.  
 
The different choices of independent momentum variables are 
the single-particle momenta $\{ \mathbf{p}_1,\mathbf{p}_2, \mathbf{p}_3\}$,
the total momentum plus the momenta of any two particles in the 
three-body rest frame,  $\{ \mathbf{P},\mathbf{k}_i, \mathbf{k}_j\}$ and
the relativistic Jacobi momenta for the $jk$ pair, 
$\{ \mathbf{P},\mathbf{k}_i, \mathbf{p}_{jk}\}$. 
The Jacobian of the 
variable change
$\{ \mathbf{p}_1,\mathbf{p}_2, \mathbf{p}_3\} \leftrightarrow
\{ \mathbf{P},\mathbf{k}_i, \mathbf{k}_j\}$ is one when 
$\mathbf{P}=0$, while the Jacobian of the variable change
$\{ \mathbf{P},\mathbf{k}_i, \mathbf{k}_j\} \leftrightarrow 
\{ \mathbf{P},\mathbf{k}_i, \mathbf{p}_{jk}\}$ is 
\beq
{\cal N}^2(\mathbf{k}_i,\mathbf{k}_j) := 
\left\vert {\partial  (\bqi,\bqj)  \over  
\partial (\bpij, \bk_{ij}) } \right \vert =
{\omega_m(p_{ij}) +\omega_m(p_{ij}) \over 
\omega_m(k_i) +\omega_m(k_j)}
{\omega_m(k_i) \omega_m(k_j) \over 
\omega_m(p_{ij}) \omega_m(p_{ij})} .
\label{2.12}
\eeq
In the limit that the momenta are much smaller than the 
masses the relativistic Jacobi momenta 
become identical to the non-relativistic Jacobi momenta
and the Jacobian becomes~1.

\subsection{Two-Body Interactions Embedded in the Three-Body Space}

Realistic two-body interactions may be e.g. motivated by meson exchange
or other effective field theories, but
the parameters of the interaction must be fine tuned in order to be
consistent with experimental two-body scattering observables. This
means that the non-relativistic interactions are already consistent
with data and at the two-body level should not be considered as
approximations to a relativistic two-body model. Instead a realistic
relativistic two-body model should be consistent with the same data.
Once the two-body model is defined, cluster properties dictate how the
two-body interactions appear in the three-body problem. This is
different in the relativistic and non-relativistic formalism.

Given a non-relativistic two-body model fit to scattering data, we
define a relativistic interaction fit to the same data by requiring
that the relativistic wave functions as a function of
$\mathbf{p}_{ij}$ are identical to the non-relativistic wave functions
as a function of the corresponding non-relativistic Jacobi momentum. 
Since the phase shifts can be extracted from
asymptotic properties of the wave functions; this ensures that both
interactions give the same phase shifts as a function of the
$\mathbf{p}_{ij}$. This can be proved using the 
invariance principle \cite{Keister:1991sb,kato:1966}. 

We begin by defining the interacting two-body invariant mass operator
(relativistic rest-frame Hamiltonian) for the $ij$ pair in terms 
of the non-relativistic two-body interaction by 
\beq
m_{jk}^2 = 4 m \biggl  ({p_{jk}^2 \over m} + v_{jk}^{nr}+m \biggr)  ,
\label{4.2}
\eeq
where $v_{jk}^{nr}$ is the non-relativistic nucleon-nucleon interaction
between particles $j$ and $k$.
Since $m_{jk}^2$  is a function of the non-relativistic rest-frame Hamiltonian, 
${p_{jk}^2 \over m} + v_{jk}^{nr}$,  
it has the same eigenfunctions. 
The phase-equivalent relativistic interaction, $v_{jk}^r$, is defined in terms of 
$m_{jk}$ by 
\beq
v_{jk}^r = m_{jk} - m_{0jk}. 
\label{4.1}
\eeq
While it is possible to formally solve the non-linear relation needed
to express $v_{jk}^r$ in terms of $v_{jk}^{nr}$~\cite{Kamada:2007ms},
this is not needed to formulate the relativistic Faddeev equation.

The input to the Faddeev equation is the two-body transition operators
properly embedded in the three-body Hilbert space. As in the 
two-body case we define interactions as the difference between the 
three-body mass operator with and without the two-body interaction.
This will satisfy $S$-matrix cluster properties if this 2+1-body mass operator
leads to the same two-body scattering operator as the two-body 
mass operator $m_{jk}^2$. This will be true if we can write the 
three-body interacting mass operator with pair $ij$ interacting as a 
function of $m_{ij}$. This can be achieved by defining
\beq
M_{jk} = \sqrt{k_i^2 + m_{jk}^2} + \omega_m (k_i )
= 
\sqrt{k_i^2 + 4m^2 + 4{p}_{jk}^2 + 
4m v^{nr}_{jk}} + \omega_m (k_i).
\label{4.4}
\eeq
The two-body interactions embedded in the three-body Hilbert space are
\begin{eqnarray} 
V_{{jk}} 
&=&
M_{jk}- M_{0}\\
&=&
 \sqrt{k_i^2 + 4 m^2 + 4 p_{jk}^2 
+ 4 m v_{jk}^{nr}}-  \sqrt{k_i^2 + 4 m^2 + 4p_{jk}^2} .
\label{4.5-2}
\end{eqnarray} 
The three body-bound states are eigenstates of the three-body mass operator
\beq
M_t= M_0 + V_{{12}} + V_{{23}} + V_{{31}}. 
\label{4.5-3}
\eeq
The Faddeev kernel involves the two-body transition operators
$T_{jk} (z)$ that act in the
three-particle Hilbert space. They are defined by
\beq
T_{jk} (z) := V_{{jk}} + V_{{jk}} (z - M_{jk})^{-1} V_{{jk}}.
\label{4.5-4}
\eeq
This is a function of the non-relativistic two-body interaction
between particles $j$ and $k$. Because of this relation, matrix
elements of $T_{jk} (z)$ can be obtained directly from the
non-relativistic two-body transition matrix elements 
using a two-step process. This
method is exact and avoids the problem of computing the
relativistic two-body interaction. 
 
The first step is to use the general relation between the interaction,
scattering wave functions and half-shell transition operators
\beq
V_{{jk}} \vert (\mathbf{k}_i, \mathbf{p}_{jk})^+ \rangle = 
T_{jk}(z_0) \vert \mathbf{k}_i, \mathbf{p}_{jk}  \rangle ,
\eeq
where $z_0 =\sqrt{m_{0jk}^2(p_{jk'})+k_i^2}+i0^+$ 
is the on-shell energy. Using the relativistic and 
non-relativistic versions of this relation leads to the identity 
\begin{eqnarray} 
T_{jk} \left (\bpjk,\bpjk';\sqrt{m_{0jk}^2(\bpjk')+\qi^2}+i0^+ \right) 
= F(\pjk,\pjk',\qi)
t_{nr} \left (\bpjk,\bpjk';\frac{p_{jk}^{\prime 2}}{m}+i0^+ \right), \cr
\label{4.6}
\end{eqnarray}
where the ratio of the half-shell transition matrix elements is
\begin{equation}
F(\pjk,\pjk',\qi)=\frac{4m}{ \sqrt{m_{0jk}^2(p_{jk}) + k_i^2} + \sqrt{m_{0jk}^2
(p_{jk}') + k_i^2} },
\label{eq.3.3-2}
\end{equation}
$m_{0jk}({p}_{jk}) = 2\omega_m ({p}_{jk})$ and 
\beq
\langle \bpjk,\bqi \vert T_{jk} (z) \vert \bpjk',\bqi' \rangle
= \delta (\bqi-\bqi' )T_{jk} \left (\bpjk,\bpjk',z- \omega_m (k_i) \right ).
\eeq
Equation (\ref{4.6}) can be used to express $T_{jk} \left
(\bpjk,\bpjk';\sqrt{m_{0jk}^2(p_{jk}')+k_i^2} +i0^+\right)$ in terms of
$t_{nr} \left (\bpjk,\bpjk';\frac{p_{jk}^{\prime 2}}{m} +i0^+ \right)$. 
The only 
problem with this relation is that it is only valid for half 
on-shell transition matrices. In the Faddeev equation 
$T_{jk} \left (\bpjk,\bpjk';z \right)$ is needed for off-shell values of $z$.

These can be obtained by solving the integral equation for
$T_{jk} \left (\bpjk,\bpjk';z \right)$
that uses Eq.~(\ref{4.6}) as input
\begin{eqnarray} 
T_{jk} \left (\bpjk,\bpjk';z- \omega_m (k_i) \right) &=& 
T_{jk} \left (\bpjk,\bpjk';\sqrt{m_{0jk}^2(p_{jk}')+k_i^2}+i0^+ \right)  \cr
&& \hskip -4.cm + 
\int d\bpjk''
\left[ {1 \over z -\sqrt{m_{0jk}^2(p_{jk}'')+k_i^2})}
- 
{1 \over \sqrt{m_{0jk}^2(p_{jk}')+k_i^2} -\sqrt{m_{0jk}^2(p_{jk}'')+k_i^2}
+i0^+}
\right ] \cr
&& \hskip -4.cm \times
T_{jk} \left (\bpjk, \bpjk'';z- \omega_m (k_i) \right) 
T_{jk} \left (\bpjk'',\bpjk';\sqrt{m_{0jk}^2(p_{jk}')+k_i^2}+i0^+ \right) . 
\label{4.7}
\end{eqnarray} 
This equation, which follows from the first resolvent equations, 
is derived in the appendix. 
Thus,  in order to compute the relativistic Faddeev kernel
one only needs to solve the integral equation, Eq.~(\ref{4.7}), 
for $T_{jk} \left (\bpjk, \bpjk'';z- \omega_m (k_i) \right)$,
which uses the non-relativistic half-on-shell transition matrix 
elements as input.

\subsection {Faddeev Equations}

The relativistic three-body bound state is a discrete eigenstate of
the relativistic three-body mass operator $M$ defined in
(\ref{4.5-3}). The eigenvalue problem can be reformulated as a system
of coupled integral equations for the Faddeev components of the wave
functions. For identical particles this reduces to a single equation
for one of the Faddeev components
\begin{equation}
|\psi_i \rangle \equiv |\psi_{jk,i}\rangle = (M_t-M_0)^{-1} \, T_{jk}(M_t)  
\, P \, |\psi_i \rangle,
\label{eq.5.1}
\end{equation}
where $M_t=E_t+3m$ is three-body mass eigenvalue, 
and $P=P_{12}P_{23} + P_{13}P_{23}$ are the standard
permutation operator for three identical particles.
This equation has non-zero solutions when $M_t$ is an eigenvalue of 
Eq.~(\ref{4.5-3}).
The bound state wave function can be constructed from
the solution of (\ref{eq.5.1}) using
\beq
\vert \Psi \rangle =
\vert \psi_i \rangle + P \vert \psi_i \rangle. 
\label{eq:5.3}
\eeq
For the explicit solution we  write them in the basis
$\vert \mathbf{k}_i , \mathbf{p}_{jk} \rangle$, where
$i$ is fixed. In this basis Eq.~(\ref{eq.5.1}) has the form
\begin{eqnarray}
\bera \bpjk , \bqi | \psi_i \ket &=& 
\int d \mathbf{p}_{jk}'  d\mathbf{k}_{i}'
d \mathbf{p}_{jk}'' d\mathbf{k}_{i}''
{  \delta (\bqi-\bqi' ) \, T_{jk} \left (\bpjk, \bpjk';M_t- \omega_m (k_i) \right )
\over M_t - M_0 ({p}_{jk} ,{k}_i)}
\langle \mathbf{p}_{jk}' ,\mathbf{k}'_i \vert P \vert \mathbf{p}_{jk}'' ,\mathbf{k}_i''
\rangle \bera \bpjk'', \bqi''| \psi_i \ket \cr  
 &=& \frac{1}{M_t-M_0(p_{jk},k_i)}
\int d \mathbf{p}_{jk}' d\mathbf{k}_{i}'
d \mathbf{p}_{jk}'' d\mathbf{k}_{i}'' \delta (\bqi-\bqi' ) \,
T_{jk} \left (\bpjk, \bpjk';M_t-\omega_m (k_i) \right) \cr 
&& \times
\biggl ( \bera \bpjk', \bqi' |  \bpki'', \bqj'' \ket   + \bera \bpjk', \bqi' |  \bpij'', \bqk'' \ket  \biggr ) \,  \bera \bpjk'', \bqi''| \psi_i \ket .
\label{eq.4.18}
\end{eqnarray} 
Here the permutation operators contain two delta functions
which eliminate two of the integrals. We use them to eliminate 
the integrals over $\mathbf{p}_{jk}'$ and  $\mathbf{p}_{jk}''$.
The matrix elements of the permutation operators have the form 
\begin{eqnarray} 
 \bera \bpjk', \bqi' | P | \bpjk'', \bqi'' \ket &=& \bera \bpjk', \bqi' | \bpki'', \bqj'' \ket  +  \bera \bpjk', \bqi' | \bpij'', \bqk'' \ket \cr
   &=& \frac{\delta^3 \biggl (\bpjk'-\bpjk(\bqi'',-\bqi'-\bqi'') \biggr) \, \delta^3 \biggl (\bpjk''-\bpjk(-\bqi'-\bqi'',\bqi') \biggr)}{{\cal N} (\bqi'',-\bqi'-\bqi'')\, {\cal N}(-\bqi'-\bqi'',\bqi')} \cr
 &&+ \frac{\delta^3 \biggl (\bpjk'-\bpjk(-\bqi'-\bqi'',\bqi'') \biggr) \, \delta^3 \biggl (\bpjk''-\bpjk(\bqi',-\bqi'-\bqi'') \biggr)}{ {\cal N}(-\bqi'-\bqi'',\bqi'')\, {\cal N}(\bqi',-\bqi'-\bqi'')},
\label{eq.4.22}
\end{eqnarray}
where ${\cal N}(\mathbf{k}_i,\mathbf{k}_j)$ is the square root of the 
Jacobian of the variable change defined in Eq.~(\ref{2.12}).

Using the symmetry property ${\cal N}(\bqj,\bqk)={\cal N}(\bqk,\bqj)$
of the Jacobian, the matrix
elements of the permutation operator $P$, Eq.~(\ref{eq.4.22}), can be written as
\begin{eqnarray} 
 \bera \bpjk', \bqi' | P | \bpjk'', \bqi'' \ket  &=& 
N(\bqi',\bqi'') \Biggl \{ \delta^3 \biggl (\bpjk'-\bpjk(\bqi'',-\bqi'-\bqi'') \biggr) \, \delta^3 \biggl (\bpjk''-\bpjk(-\bqi'-\bqi'',\bqi') \biggr)  \cr
&& \hskip 2cm + \delta^3 \biggl (\bpjk'-\bpjk(-\bqi'-\bqi'',\bqi'') \biggr) \, \delta^3 \biggl (\bpjk''-\bpjk(\bqi',-\bqi'-\bqi'') \biggr)  \Biggr \}, \nonumber \\
\label{eq.4.24}
\end{eqnarray}
where
\beq 
N(\bqi',\bqi'') = {\cal N}^{-1}(-\bqi'-\bqi'',\bqi'') \, {\cal N}^{-1}(-\bqi'-\bqi'',\bqi').
\label{eq.4.25}
\eeq
Inserting Eq.~(\ref{eq.4.24}) into Eq.~(\ref{eq.4.18})
and integrating over the delta functions leads to the
relativistic Faddeev integral equation
\begin{eqnarray} 
\langle \bpjk \,, \bqi \vert \psi_i \rangle =
 \frac{1}{M_t-M_0(p_{jk},k_i)}\int d \bqi' \, N(\bqi,\bqi') \,  T_{jk}^{sym} 
\biggl(\bpjk,\btpi;M_t-\omega_m (k_i) \biggr)  \, 
\langle \bpi,\bqi'\vert \psi_i \rangle, \cr 
\label{eq.4.26}
\end{eqnarray}
where $T_{jk}^{sym}$ is the symmetrized boosted two-body $T$-matrix, 
defined by 
\begin{eqnarray} 
 T_{jk}^{sym} \bigl ( \bpjk,\bpjk';\epsilon \bigr) 
=   T_{jk} \bigl ( \bpjk,\bpjk';\epsilon \bigr)  +  T_{jk}\bigl ( -\bpjk,\bpjk';\epsilon \bigr) .
\label{eq.4.26-2}
\end{eqnarray}
and 
\begin{eqnarray} 
\btpi &=& \bpjk(\bqi',-\bqi-\bqi') = \bqi' + \frac{1}{2} C(\bqi,\bqi') \, \bqi  , \cr
\bpi &=& \bpjk(\bqi+\bqi',-\bqi') =  \bqi  + \frac{1}{2} C(\bqi',\bqi) \, \bqi'.
\label{eq.4.28}
\end{eqnarray}
The coefficient $C(\bqi,\bqi')$ is defined as \cite{Lin:2007ck}
\begin{eqnarray} 
C(\bqi,\bqi') \equiv 1+ \frac{ \omega_m (k_i')- 
\omega_m(\vert\bqi + \bqi'\vert) }
{ \omega_m(k_i') + \omega_m(\vert \bqi + \bqi'\vert ) + 
\sqrt{\biggl ( \omega_m(k_i') 
+ \omega_m (\vert\bqi + \bqi' \vert) \biggr)^2-k_i^2} }.
\label{eq.4.29}
\end{eqnarray}
In deriving  Eq.~(\ref{eq.4.26}) we used the property
\begin{eqnarray} 
 \bera \bpjk , \bqi | \psi_i \ket
&=& \bera -\bpjk , \bqi | \psi_i ,\ket. 
\label{eq.4.27}
\end{eqnarray}
The relativistic momenta $\btpi$ and $\bpi$ defined in
Eq.~(\ref{eq.4.28}) become the corresponding non-relativistic ones if
the coefficient $C(\bqi,\bqi')$ is equal to one, which is the case if
the momenta are small with respect to the masses.

To solve the integral equation Eq.~(\ref{eq.4.26}), 
we follow Ref.~\cite{Elster:1998qv} and  choose a 
coordinate system where $\bqi$ is parallel to $z-$axis and $\bpjk$ 
is in the $x-z$ 
plane. Then the variables that appear in the Faddeev integral equation 
are magnitudes of the vectors as well as angles between them.
They are
\begin{eqnarray}
  x &\equiv& x_{\pjk} = \hat{\bq}_i \cdot \hat{\bp}_{jk}, \cr
x'  &\equiv& x_{\qi'} = \hat{\bq}_i \cdot \hat{\bq}_i'  , \cr
y &\equiv& x_{\pjk \qi'} = \hat{\bp}_{jk} \cdot \hat{\bq}_i' = xx'+\sqrt{1-x^{2}}\sqrt{1-x'^{2}}\cos (\phi_{\qi'} ),
 \cr
\tpi&=&\sqrt{\frac{1}{4}C^2(\qi,\qi',x_{q'}) \qi^{2}+\qi'^{2}+ C(\qi,\qi',x_{\qi'}) \qi \qi' x_{\qi'}} ,
  \cr 
  \pi&=&\sqrt{\qi^{2}+\frac{1}{4} C^2(\qi',\qi,x_{q'}) \qi'^{2}+C(\qi',\qi,x_{\qi'}) \qi \qi' x_{\qi'}},
    \cr
 x_{\pjk \tpi}&=&\frac{\frac{1}{2}C(\qi,\qi',x_{\qi'})\qi x_{\pjk}+\qi' y}{\tpi},
 \cr
x_{\pi \qi'} &=& \frac{\qi x_{\qi'} +\frac{1}{2} C(\qi',\qi,x_{\qi'}) \qi'}{\pi} .
\label{eq.4.29-2}
 \end{eqnarray}
Using these variables, Eq.~(\ref{eq.4.26}) takes the explicit form
\begin{eqnarray}
\langle \pjk,\qi,x_{\pjk} \vert \psi_i \rangle  &=& \frac{1}{M_t-  
\omega_m (k_i) - \sqrt{m_{0jk}^2(\pjk)+k_i^2} } \,  \int_{0}^{\infty} d\qi' \qi'^2
\int_{-1}^{+1} dx_{\qi'} \int_{0}^{2\pi} d\phi_{\qi'}   
\cr &\times&
 N(\qi,\qi',x_{\qi'}) \, T^{sym}_{{jk}} \biggl ( \pjk,\tpi, x_{\pjk \tpi} ; M_t-
\omega_m (k_i) \biggr)
\langle \pi,\qi', x_{\pi \qi'} \vert \psi_i \rangle .
 \label{eq.4.30}
 \end{eqnarray}
To solve Eq.~(\ref{eq.4.30}), which has the form $x = K(M_t) x$,
we treat it as eigenvalue problem of the form 
$\lambda x = K(M_t) x$ and vary $M_t$ until $\lambda$ is 1 to a given 
precision.

\subsection{Three-Body Wave Function}

Once the Faddeev component $\langle \bpjk,\bqi \vert \psi_i \rangle$
is calculated, the three-body wave function can be obtained from
Eq.~(\ref{eq:5.3}) as
\begin{equation}
\langle \bpjk,\bqi \vert \Psi \rangle  
=\langle \bpjk,\bqi \vert \psi_i \rangle  + 
\left | \frac{\partial(\bpki,\bqj)}{\partial(\bpjk,\bqi)}  \right|^{1/2}   
\langle \bpki,\bqj \vert \psi_i \rangle  +  \left | \frac{\partial(\bpij,\bqk)}{\partial(\bpjk,\bqi)}  \right|^{1/2} \langle \bpij,\bqk \vert \psi_i \rangle,
 \label{eq.4.31}
\end{equation}
where the Jacobi momenta in systems $(ki,j)$ and $(ij,k)$ are connected to the ones in
system $(jk,i)$, i.e. $\bpjk,\bqi$, by 
 \begin{eqnarray}
 \bqj &=& \bpjk + \frac{\bqi}{m_{0jk} }  \left ( \frac{\bqi \cdot \bpjk}{m_{0jk} +\sqrt{m_{0jk}^2 +k_i^2}}   - \omega_m (p_{jk} )   \right)  ,\cr
 \bqk &=& -\bqi - \bqj =
-\bqi  - \bpjk - \frac{\bqi}{m_{0jk} }  \left ( \frac{\bqi \cdot\bpjk}{m_{0jk} +\sqrt{m_{0jk}^2 +k_i^2}}   - \omega_m (p_{jk} )   \right)  ,\cr
 \bpij &=& \bqi + \frac{\bqk}{m_{0ij} }  \left ( \frac{\bqk \cdot\bqi}{m_{0ij} +\sqrt{m_{0ij}^2 +k_k^2}}   + \omega_m (k_i)  \right) , \cr
 \bpki &=& \bqk + \frac{\bqj}{m_{0ki} }  \left ( \frac{\bqj \cdot\bqk}{m_{0ki} +\sqrt{m_{0ki}^2 +k_j^2}}   + \omega_m (k_k)  \right) ,
  \label{eq.4.32}
\end{eqnarray}
with
 \begin{eqnarray}
m_{0ij}  &=&  \sqrt{ ( \omega_m (k_i) + \omega_m(k_j)  )^2- k_k^2} , \cr
m_{0ki}  &=&  \sqrt{ ( \omega_m (k_k) + \omega_m (k_i) )^2- k_i^2} .
  \label{eq.4.33}
\end{eqnarray}

The Jacobians for changing the basis states from system $(jk,i)$ to ($ki,j$) and $(ij,k)$ are
 \begin{eqnarray}
\left | \frac{\partial(\bpjk,\bqi)}{\partial(\bpki,\bqj)}  \right| &=&
\left | \frac{\partial(\bpjk,\bqi)}{\partial(\bqj,\bqk)}  \right| 
\left | \frac{\partial(\bqj,\bqk)}{\partial(\bqi,\bqk)}  \right|
\left | \frac{\partial(\bqi,\bqk)}{\partial(\bpki,\bqj)}  \right| 
= \frac{{\cal N}^{2}(\bki,\bkk)}{{\cal N}^{2}(\bkj,\bkk)}   ,
\cr 
\cr 
\left | \frac{\partial(\bpjk,\bqi)}{\partial(\bpij,\bqk)}  \right| &=&
\left | \frac{\partial(\bpjk,\bqi)}{\partial(\bqj,\bqk)}  \right| 
\left | \frac{\partial(\bqj,\bqk)}{\partial(\bqi,\bqj)}  \right|
\left | \frac{\partial(\bqi,\bqj)}{\partial(\bpij,\bqk)}  \right| 
=  \frac{ {\cal N}^{2}(\bki,\bkj)}{{\cal N}^{2}(\bkj,\bkk)} .
  \label{eq.4.33-2}
\end{eqnarray}
The Jacobi momenta in Eq.~(\ref{eq.4.32}) can be explicitly given as a function of the Jacobi momenta of system $(jk,i)$,
 \begin{eqnarray}
 \bqj &=& \bpjk + \alpha \bqi ,\cr
 \bqk &=& -\bpjk - \beta \bqi , \cr
 \bpij &=& \gamma_p \bpjk +  \gamma_k \bqi, \cr
 \bpki &=& \xi_p \bpjk +  \xi_k \bqi,
  \label{eq.4.44}
\end{eqnarray}
where
 \begin{eqnarray}
  \alpha &=& \frac{1}{m_{0jk}} \left (  \frac{\pjk \qi x_{\pjk}  }{ m_{0jk} + \sqrt{ m_{0jk}^2 + \qi^2} } - \frac{1}{2} m_{0jk}  \right), \cr
  \beta &=& 1+ \alpha , \cr
  \gamma_p &=& \frac{1}{m_{0ij}} \left (  \frac{ \pjk \qi x_{\pjk} + \beta \qi^2  }{ m_{0ij} + \sqrt{ m_{0ij}^2 + \qk^2} } + m_{0i}  \right), \cr 
  \gamma_k &=&    1+ \gamma_p  \beta , \cr
  \xi_p &=&   -1 - \frac{1}{m_{0ki}} \left (  \frac{ \pjk^2 + \alpha \beta \qi^2 + (\alpha+\beta) \pjk \qi x_{\pjk}  }{ m_{0ki} + \sqrt{ m_{0ki}^2 + \qj^2} } -  m_{0k}  \right), \cr
  \xi_k &=&   \alpha ( \xi_p+1 )  -\beta .
  \label{eq.4.45}
\end{eqnarray}
In the coordinate system defined by Eq.~(\ref{eq.4.29-2}), the relativistic
three-body wave function of Eq.~(\ref{eq.4.31}) has the form
\begin{equation}
\langle \pjk,\qi,x_{\pjk}\vert \Psi \rangle =
\langle \pjk,\qi,x_{\pjk} \vert \psi_i \rangle  +  
\frac{{\cal N}(\bkj,\bkk)}{{\cal N}(\bki,\bkk)}  
\langle \pki,\qj,x_{\pki \qj} \vert \psi_i \rangle  + 
\frac{{\cal N}(\bkj,\bkk)}{{\cal N}(\bki,\bkj)} 
\langle \pij,\qk,x_{\pij \qk} \vert \psi_i \rangle ,
 \label{eq.4.46}
\end{equation}
with
\begin{eqnarray}
 \pki &=& | \bpki |  = \left | \xi_p \bpjk +  \xi_k \bqi \right | = \sqrt { \xi_p^2 \, \pjk^2 + \xi_k^2 \, \qi^2 + 2\, \xi_p \, \xi_k\, \pjk \,\qi \, x_{\pjk} }, 
 \cr
\qj &=& | \bqj | = \left |  \bpjk + \alpha \bqi  \right |  =  \sqrt{  \pjk^2 +\alpha^2 \qi^2 +2 \, \alpha \, \pjk \, \qi \, x_{pjk}  }, \cr
x_{\pki \qj} &\equiv& \hat{\bp}_{ki} \cdot \hat{\bq}_j =   \frac{ \xi_p\, \pjk^2 + \alpha \, \xi_k \, \qi^2  + (\alpha \, \xi_p + \xi_k) \pjk \, \qi \, x_{pij} }{ \pki \, \qj}, 
\cr
\pij &=& | \bpki |  = \left | \gamma_p \bpjk +  \gamma_k \bqi \right | = \sqrt { \gamma_p^2 \, \pjk^2 + \gamma_k^2 \, \qi^2 + 2\, \gamma_p \, \gamma_k\, \pjk \,\qi \, x_{\pjk} }, 
 \cr
\qk &=& | \bqk | = \left | - \bpjk - \beta \bqi  \right |  =  \sqrt{  \pjk^2 +\beta^2 \qi^2 +2 \, \beta \, \pjk \, \qi \, x_{pjk}  }, \cr
x_{\pij \qk} &\equiv& \hat{\bp}_{ki}   \cdot \hat{\bq}_j = -  \frac{ \gamma_p\, \pjk^2 + \beta \, \gamma_k \, \qi^2  + (\beta \, \gamma_p + \gamma_k) \pjk \, \qi \, x_{pjk} }{ \pij \, \qk}.
 \label{eq.4.47}
\end{eqnarray}


\section{Results and Discussion}
\label{results}

\subsection{Binding Energy}

To evaluate the relativistic effects in the three-body binding energy we use
the Malfliet-Tjon V~\cite{Malfliet:1968tj} potential, 
\begin{eqnarray}
\langle \mathbf{p}_{ij} \vert v^{nr} \vert \mathbf{p}_{ij}' \rangle 
=\frac{1}{2\pi^2}\left( \frac{V_R}{
(\mathbf{p_{ij}}- \mathbf{p_{ij}}^\prime)^2    +\mu_R^2} - \frac{V_A}
{( \mathbf{p_{ij}}- \mathbf{p_{ij}}^\prime)^2   +\mu_A^2} \right).
 \label{eq.3.3}
\end{eqnarray}
with parameters given in
in Ref. \cite{Elster:1998qv}. In addition, we use
$\hbar^2/m=41.470$~MeV\,fm$^2$ and $\hbar c = 197.3286$~MeV\,fm. 
In order to numerically solve 
Eq.~(\ref{eq.4.30}), the integrals over
the continuous momenta and angle variables are replaced by sums
over discrete quadrature points.
To reach five significant digit
convergence in binding energy we use
100 Gaussian quadrature points for the Jacobi momentum $\pjk$ on the interval 
[0, 60 fm$^{-1}$], 60 quadrature points for
Jacobi momentum $\qi$ on the interval [0,20 fm$^{-1}$], and  
40 quadrature points for the angle variables. 
The Faddeev integral equation, Eq.~(\ref{eq.4.30}), is solved
by iteration using a Lanczos algorithm
\cite{Stadler:1991zz}. 
The iteration of this integral equation requires a large number
of two-dimensional interpolations on the Faddeev component and
symmetrized two-body $t-$matrix. We performed the interpolation
using cubic-Hermite splines of Ref.~\cite{Huber:1996td}. 
Seven iterations are sufficient to search for the  mass eigenvalue
with a relative error of $10^{-6}$.

\begin{figure}
\begin{center}
\includegraphics[width=14cm]{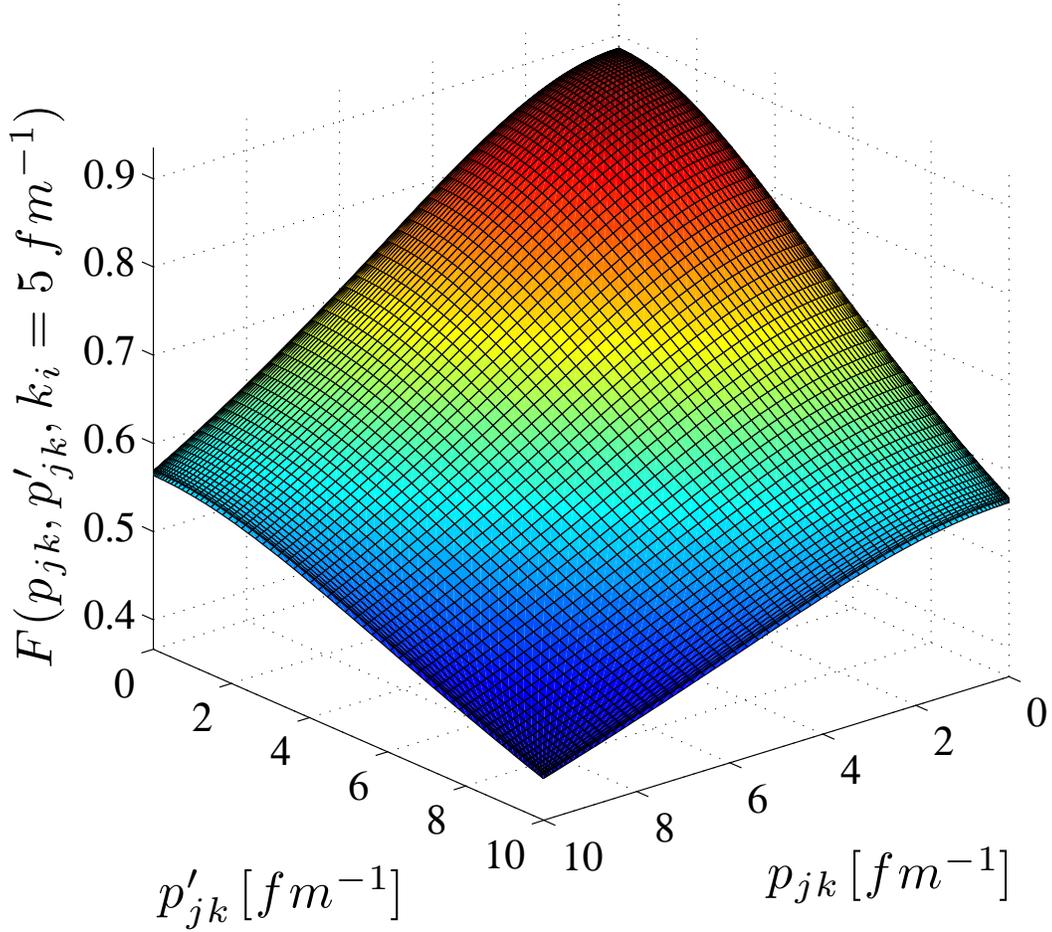}
\end{center}
\caption{(Color online) The  analytical term $F(\pjk,\pjk',\qi)$ connecting
the relativistic and non-relativistic right-half-shell $T$-matrices,
Eq.~(\ref{eq.3.3-2}), as function of the   momenta $\pjk$ and $\pjk'$ in the
two-body subsystem, for a fixed third particle momentum $\qi=5\,\text{fm}^{-1}$.}
\label{fig1}
\end{figure}

The off-shell $T$-matrix, which is needed as input for the Faddeev
integral equation, is computed by solving 
Eq.~(\ref{eq.a8}). The input is the
right-half-shell $T$-matrix embedded in the three-body Hilbert space,
which is analytically obtained from the non-relativistic $T$-matrix by
Eq.~(\ref{4.6}). In Fig.~\ref{fig1} we plot the ratio 
$F(\pjk,\pjk',\qi)$ defined in Eq.~(\ref{eq.3.3-2})
of these two transition operators
as a function of momenta $\pjk$ and $\pjk'$ for third particle momentum $\qi=5\, \text{fm}^{-1}$.
The slope of this function decreases as the value of $\qi$ increases. 
 
As a numerical test of the solution of first resolvent integral
equation for negative energies, we reproduce the same non-relativistic
three-body  binding energy as obtained from the direct solution of the
Lippmann-Schwinger (LS) equation for the off-shell $T$-matrix within four
significant figures. The imaginary part of the transition
matrix calculated from the first resolvent equation is of the
order of $10^{-11}$ MeV$^{-2}$, which is $10^4$ times smaller than the
real part,  which gives an additional measure of the accuracy of 
the calculation.

In order to graphically
analyze the Jacobian function $N(\qi,\qi',x_{\qi'})$, which appears
directly in the kernel of Faddeev integral equation, 
we parameterize it as $N(k \cos (\theta),k \sin (\theta), x \equiv x_{\qi'})$,
and plot it as a function of $\theta$ and $x$ for 
$k=1, \, 5,\, 10$ and $20$ fm$^{-1}$.
We use the same representation for the
matrix elements of permutation coefficient
$C(k\cos(\theta),k\sin(\theta),x \equiv x_{\qi'})$ shown in
Fig~\ref{fig3}. 

The solution of the relativistic Faddeev equation leads to the
three-body binding energy, $E_t^{r}=-7.4825$~MeV, which is
slightly less than the non-relativistic 
binding energy of $E_t^{nr}=-7.7382$~MeV. 
Thus, the relativistic effect is small, about 3.3\%.  
This is consistent with a reduction of 2.7\% for
the $s-$wave calculation  of Gl\"{o}ckle et
al. \cite{Glockle:1986zz}. 

The difference between the relativistic and non-relativistic
calculations come from (1) the Jacobian function $N$, (2) the
permutation coefficient $C$, (3) the relation between the relativistic
and non-relativistic $T_{ij}(z)$ and (4) the relations between the 
relativistic and non-relativistic free Green functions. 
If we keep leading non-zero terms in the limit that the
masses are larger than the momenta, all four of these factors reduce to
the corresponding non-relativistic quantities. For the off-shell
transition matrix, the kernel of the first resolvent equation reduces
to the non-relativistic kernel in the same limit, which implies that
both the half-shell and off-shell two-body transition matrix elements approach
their non-relativistic counterparts. The fact that all four corrections 
become small in that limit 
suggests that momentum/mass expansions are valid approximations.
However, the corrections that relate the relativistic and non-relativistic 
Faddeev equation are only due to relativistic effects associated with a
relativistic treatment of the Fermi motion with respect to the spectator 
particle. It would be incorrect to apply these expansions to the 
two-body dynamics,  which are fit to the same data in both the relativistic 
and non-relativistic case. 

While our calculations suggest that relativistic effects are small,
it is important to remember is that only the combination of all
four ingredients 
 leads to a small correction, while individually 
the  corrections do not have to be small.
In Table~\ref{table1}  the
contributions of the Jacobian function $N$ and the permutation
coefficient $C$  to the relativistic three-body binding energy are shown.
Both of these functions become~1 in the non-relativistic limit.
By setting the Jacobian function $N$ to 1 in our relativistic
calculations, the binding energy increases 
about 0.8 \%. 
Setting the permutation coefficient $C$ to 1 
leads to a small decrease, about 0.6\% of
the  binding energy. Finally, by setting both, the
Jacobian function $N$ and the permutation coefficient $C$ to 1,
the three-body binding energy has a small
increase of about 0.15\%. This means that ignoring the Jacobian function $N$
and the permutation coefficient $C$ in the relativistic formalism,
leads to less than 0.2\% over-binding. The combined effect
of the Jacobian and permutation operators is a factor of 4-5 smaller
than the effect of each one individually. The main contribution of
relativistic effects in the three-body binding energy comes from the
relativistic transition operator and free propagator. To evaluate the
contribution of relativistic $T$-matrix to the three-body  
binding energy, we
replace the relativistic $T$-matrix by non-relativistic one
in the kernel of relativistic Faddeev integral equation. This 
substitution results in an
increase of about 2.4\% in the energy. When replacing the 
relativistic free propagator the by non-relativistic one, a decrease
of 1.8\% in relativistic 3B binding energy can be observed. 
These numerical results imply that the main contribution of relativistic effects in three-body binding
energy results from the relativistic $T$-matrix. 
The remaining contribution stem from the free propagator, the Jacobian
function and the permutation coefficient.

\subsection{Three-body Wave Function and Momentum Distribution}

Using the Faddeev component from Eq. (\ref{eq.4.30})
the total wave function $\Psi(\pjk,\qi,x_{\pjk})$ can be obtained
by  three-dimensional interpolations on momentum and angle variables, 
as shown in Eq. (\ref{eq.4.46}). The wave function is normalized as
\begin{equation}
\bera \Psi  | \Psi\ket = 
8 \pi^2 \int_0^{\infty} d\pjk \, \pjk^2 
\int_0^{\infty} d\qi \, \qi^2 \int_{-1}^{+1} dx_{\pjk} \,
\Psi^2(\pjk,\qi,x_{\pjk}) = 1.
\label{eq.3.4}
\end{equation}
The left panel of Fig. \ref{fig4} shows contour plots of the
logarithm of the absolute value of the relativistic total wave functions for
fixed angle $x_{\pjk}=0$ (top left) and $x_{\pjk}=+1$ (bottom
left). The right panel shows the difference between the relativistic
and corresponding non-relativistic wave functions. These figures
indicate
that the largest relativistic effect appear at  large values of the
momentum of third particle $\qi$. This is not surprising, since 
the primary relativistic effects are expected to be due to the Fermi motion.

In order to simplify our analysis of the three-body 
wave function and to provide insight on the structure of wave
function, we calculate the momentum distribution function $n(\qi)$,
the probability to find a particle with momentum $\qi$ in the nucleus,
and $n(\pjk)$, the probability to find a pair with momentum $\pjk$ in
the nucleus, which are defined as
\begin{eqnarray}
n(\pjk) &=& 8 \pi^2 \pjk^2 \int_0^{\infty} d\qi \qi^2 \int_{-1}^{+1} dx_{\pjk} \,
\Psi^2(\pjk,\qi,x_{\pjk}) , \cr
n(\qi) &=& 8 \pi^2 \qi^2 \int_0^{\infty} d\pjk \pjk^2 \int_{-1}^{+1} dx_{\pjk} \,
\Psi^2(\pjk,\qi,x_{\pjk}) .
 \label{eq.3.5}
 \end{eqnarray}
Electron scattering is sensitive to 
the quantity $n(\qi)$. By considering the normalization of total wave
function, given in Eq. (\ref{eq.3.4}), both momentum distribution
functions are also normalized to one, i.e. $\int_0^{\infty} n(\pjk)
d\pjk = 1$ and $\int_0^{\infty} n(\qi) d\qi = 1$. The momentum
distribution functions $n(\pjk)$ and $n(\qi)$ calculated from
relativistic and non-relativistic wave functions are presented in
Fig. \ref{fig5}. The difference between relativistic and
non-relativistic momentum distribution functions appears to be very
small, but as we have shown in Fig. \ref{fig6}, the
differences for both $n(\pjk)$ and $n(\qi)$ have a peak at $\pjk \sim
0.2\, fm^{-1}$ and $\qi \sim 0.2\, fm^{-1}$ and have a dip at $\pjk
\sim 0.7\, fm^{-1}$ and $\qi \sim 0.7\, fm^{-1}$. 
The peak of $\Delta
n(\pjk)=n_r(\pjk)-n_{nr}(\pjk)$ and $\Delta n(\qi)=n_r(\qi)-n_{nr}(\qi)$ 
has a shift to up for setting the permutation coefficient $C$ to one, whereas
by setting the Jacobian function $N$ to one, it has a shift to down and by setting both
Jacobian and permutation coefficient to one, the result is a small
shift to down. The behavior for the dip is reversed.

\section{Summary and Outlook} 

In this work we solved the relativistic momentum-space Faddeev equation for
three nucleons interacting with a spinless Malfliet-Tjon type potential without
partial-wave decomposition for the three-body binding energy and calculated the
corresponding bound-state wave function.
In order to identify relativistic effects the
relativistic two-body interaction was defined so it gives in the
two-body rest frame the same
phase shifts and wave functions as the non-relativistic interaction.
Relativistic effects arise since the interacting two-body subsystems are not at
rest in the three-body rest frame.  Lorentz boosts
associated with each subsystem are used to transform the two-body
interactions from the two-body rest frames to the three-body
rest frame.  This transformation is determined by 
both the relativistic symmetry and cluster properties.

The requirement that the relativistic and non-relativistic two-body
interactions be phase equivalent is due to the fact that realistic
non-relativistic nucleon-nucleon interactions are already designed to
be consistent with experiment, and thus are consistent with special
relativity.  This means that all of the observable effects of special
relativity are related to the different ways that the relativistic
and non-relativistic problems treat the Fermi motion.  The relativistic
Faddeev equation is simply a reformulation of the eigenvalue problem
for the relativistic mass operator. However, it has the important
advantage that in the limit that the mass scales are large
compared to the momentum scales, both the variables and the kernel of the
integral equation
approach the corresponding quantities that appear in the
non-relativistic Faddeev equation.  This relation is not as transparent
when one compares the relativistic mass operator and the
non-relativistic rest Hamiltonian.  This also suggests that as long as
the Fermi momentum scales are small compared to the mass scales
relativistic corrections are expected to be small.  Of course,
this naive picture is impacted by exchange symmetry, and because
the Malfliet-Tjon potential is a relatively hard potential, i.e. high momenta
being involved, it needs to be verified by a calculation.

A comparison of the relativistic and non-relativistic equations show
four essential differences. Two are related to the
Jacobi momenta, which leads to different treatments of the permutation
operator in the relativistic and non-relativistic Faddeev kernels.
While these are choices of variables, the relativistic mass operator
(\ref{4.5-3}) that has an $S$-matrix that clusters is naturally
expressed in terms of the relativistic Jacobi momenta.  
The difference
between these variables and the corresponding non-relativistic Jacobi
momenta appear in the coefficients C defined in Eq.~(\ref{eq.4.29}) and
the non-trivial Jacobians (factors $N$  of Eq.~(\ref{eq.4.25})) 
in the relativistic case. 
For $C\to 1$ and $N \to 1$ these become the non-relativistic Jacobi momenta.  
The combination of these two factors are associated with the
kinematics of cluster properties.  
An important observation of our calculations is 
that those two quantities have opposite effects on the value of the binding
energy, and that the combination of
those to quantities is essential to have a net effect, which is about a factor
of four to five smaller than each effect separately. 

The other two areas where relativity
plays a role is in the part of the Faddeev kernel involving the 
two-body transition operator and the free three-body Green's functions.  
The difference between the relativistic and non-relativistic  
half-shell $T$ is contained in the ratio $F$, Eq.~(\ref{eq.3.3-2}), which becomes
1 in the limit that the masses are much larger than the momentum 
scales.  This is input to the kernel of the first resolvent equation 
so there are similar correction to the off shell transition 
matrix elements. The combination $T(z) g_0(z)$ is dimensionless,
in both the relativistic and non-relativistic case.  Again,
setting $F$  to one without making a corresponding change in the 
free Green's function results in a significant increase in the binding energy.
Replacing the relativistic propagator by the 
non-relativistic one and keeping the function $F$, the binding energy decreases 
roughly the same amount. The combination of those two effects leads to a
relatively small increase in the binding energy.

We calculated the relativistic three-body wave function and compared it to its
non-relativistic counterpart. 
The largest difference can be seen in the dependence on the
spectator momentum $k_i$. This is not surprising since the
$k_i$ dependence is dictated by the different ways that the relativistic and 
non-relativistic calculations treat the Fermi motion.   
Finally, we calculate that relativistic effects decrease the binding 
energy by about 3.3\%.  This is consistent with the results of
the calculations of \cite{Glockle:1986zz,Kamada:2008xc}.

This work demonstrates that direct 
integration techniques can be used to achieve the same results 
that can be obtained using partial-wave methods.  Since our
long-term interest is to first replace the Malfliet-Tjon interaction by
a realistic interaction that has a more complicated spin-isospin 
dependence, and then to extend the calculations to treat scattering 
at the few GeV scale, it is important to test these methods in successive
steps.


\appendix

\section{The Boosted Off-Shell $T$-matrix Obtained via  Resolvent Equations}
\label{appendixA}

\subsection{Resolvent Equations}

Starting from the resolvent of $M$, $g(z_i) = (z_i - M)^{-1}$, where $z_i = E_i
+i\epsilon$ one obtains the first resolvent equation that relates the 
resolvent at
two different values of $z_i$ as
\begin{eqnarray}
g(z_j) &=& g(z_i) + \bigl[ g(z_j) - g(z_i)\bigr] \cr
     &=&  g(z_i) + g(z_j) \left[g^{-1}(z_i) - g^{-1}(z_j) \right] g(z_i).
\label{eq:a1}
\end{eqnarray} 
Multiplying Eq.~(\ref{eq:a1}) from the left and right by an interaction 
operator $V$ and adding $V$ to both sides of the equation leads to
\begin{eqnarray}
V+Vg(z_j)V &=& V+Vg(z_i)V +V \left[ g(z_j) - g(z_i)\right] V ,
\label{eq:a2}
\end{eqnarray}
or
\begin{eqnarray}
 T(z_j)   &=& T(z_i)    + V g(z_j) \left[ z_i -z_j\right] g(z_i) V \cr
  &=& T(z_i)     + T(z_j) g_0 (z_j) \left[ z_i -z_j\right] g_0(z_i)  T(z_i) \cr 
   &=& T(z_i)  + T(z_j) \left[ g_0 (z_j) - g_0(z_i)\right] T(z_i),
\label{eq:a2-2}
\end{eqnarray}
where  we used the identity $\frac{1}{A} - \frac{1}{B} = \frac{1}{A} (B-A)
\frac{1}{B}$, the definition of the transition operator $T(z_i) =V+  Vg(z_i)V$
 as well as the identity $ g(z_i)V \equiv  g_0(z_i) T(z_i)$, with $g_0(z_i)
= (z_i -M_0)^{-1}$ being the resolvent of $M_0$. We now obtain an integral relation,
which connects the transition operator at a given energy argument $z_i$ with the transition operator at a
different energy $z_j$. 

Next, we take matrix elements, $\langle {\bf p}|T(z_i)|{\bf p'}\rangle 
\equiv T({\bf p},{\bf p'};z_i)$. Since the relativistic $T$-matrix is only 
known for
half-shell momentum variables, i.e. $T({\bf p}, {\bf p_i}; z_i)$, we need to take the matrix elements
of Eq.~(\ref{eq:a2-2}), which leads to the inhomogeneous integral equation
\begin{equation}
\langle {\bf p}|T(z_j)|{\bf p_i}\rangle = \langle {\bf p}|T(z_i)|{\bf p_i}\rangle 
 + \int d^3{\bf p''} \langle {\bf p}|T(z_j)|{\bf p''}\rangle \left[ g_0 (z_j) -
g_0(z_i)\right] \langle {\bf p''}| T(z_i)|{\bf p_i}\rangle .
\label{eq:a3}
\end{equation}
Here the inhomogeneous term is given by the half-shell $T$-matrix elements
$\langle {\bf p}|T(z_i)|{\bf p_i}\rangle = T({\bf p},{\bf p_i};z_i)$, and we solve
for the off-shell matrix elements 
$\langle {\bf p}|T(z_j)|{\bf p_i}\rangle = T({\bf p},{\bf p_i}; z_j)$.

\subsection{Numerical Realization}

Writing Eq.~(\ref{eq:a3}) explicitly using $z_i \equiv E(p_i) $ leads
to 
\begin{eqnarray}
T({\bf p},{\bf p_i};E(p_j)) &=& T({\bf p},{\bf p_i};E(p_i)) \cr
&& \hskip -2cm +   \int d^3{\bf p''} 
T({\bf p},{\bf p''}, E(p_j)) \left[\frac{1}{E(p_j) - E(p'')+i\epsilon} -
\frac{1}{E(p_i)- E(p'')+i\epsilon}\right] T({\bf p''},{\bf p_i};E(p_i)). \nonumber \\
\label{eq:a4}
\end{eqnarray}
Choosing the vector ${\bf p}$ parallel to the z-axis and the vector ${\bf p_i}$ in
the x-z plane leads to the following angle variables
\begin{eqnarray}
\hat{\mathbf{p}}\cdot\hat{\mathbf{p}}_i &\equiv & x_i \cr
\hat{\mathbf{p}}'' \cdot\hat{\mathbf{p}}_i &\equiv & \hat{x}_i 
= x_i x''
-\sqrt{1-x_i^2}\sqrt{1-x''^2} \cos \phi_{p''p_i} \equiv {\hat x}_i(x_i,x'',\cos
\phi_{p''p_i})\cr
{\hat p}\cdot {\hat p}'' &\equiv & x'' \; .
\label{eq:a5}
\end{eqnarray}
Inserting the above variables into  Eq.~(\ref{eq:a5}) leads to
\begin{eqnarray}
T(p,p_i,x_i;E(p_j)) &=& T(p, p_i,x_i;E(p_i))  \cr
&+& \int_0^\infty dp \; p'' \int_{-1}^{+1} dx'' 
\left[ \frac{1}{E(p_j)-E(p'')+i\epsilon}-\frac{1}{E(p_i)-
 E(p'')+i\epsilon}\right]  \cr
 &\times&
 T(p,p'',x''; ;E(p_j)) 
\int_0^{2\pi} d\phi'' T(p'',p_i,{\hat x}_i(x_i,x'',\cos \phi_{p''p_i});E(p_i)).
\label{eq:a6}
\end{eqnarray}
Here we note that the $\phi''$ integration only affects one term in the integral
equation and we can carry it out separately. 
For convenience let us define
\begin{equation}
T(p'',p_i,x_i,x'';E(p_i))\equiv \int_0^{2\pi} d\phi'' T(p'',p_i,{\hat
x}_i(x_i,x'',\cos \phi_{p''p_i});E(p_i)).
\label{eq:a7}
\end{equation}
The structure of Eq.~(\ref{eq:a6}) is identical to the two-body LS 
equation~\cite{Elster:1997hp} and can be solved in a similar fashion.
However, one needs to carefully look at its singularities. For the calculation of
the relativistic three-body bound state equation, we need the off-shell $T$-matrix at
negative energies $E(p_j)$. Thus, the first propagator in Eq.~(\ref{eq:a6}) is
non-singular, and its numerical value always negative. However, the second
propagator exhibits a singularity at $E(p_i)=E(p'')$ for each fixed momentum $p_i$. 
This singular point on the momentum grid can be numerically treated with a
subtraction technique.
In the actual calculation, we work on a momentum grid for $p_i$, and we use the
same momentum grid for the integration over $p''$. In this case, when setting up the
matrix equation to solve Eq.~(\ref{eq:a6}), the singular point is located on the
diagonal of this matrix, and all terms resulting from the analytic treatment of the
singularity must be located on the diagonal.
 
In order to calculate the off-shell boosted $T$-matrix $T_{\qi} \left
  (\pjk,\pjk',x_{\pjk'}; M_t-m_{0i}(k_i) \right)$ which appears in the
kernel of Faddeev integral equation from right-half-shell boosted $T$-matrix $T_{\qi} \left
  (\pjk,\pjk',x_{\pjk'};\sqrt{m_{0jk}^2(p{jk}')+k_i^2} \right)$ we
solve Eq.~(\ref{eq:a6}) for $E(p_j)=M_t-m_{0i}(k_i)$ and
$E(p_i)=\sqrt{m_{0jk}^2(p_{jk}')+k_i^2}$.

After calculating the singularity of the second integral of
Eq.~(\ref{eq:a6}) with the  subtraction method, the explicit form of the first resolvent integral equation is obtained as
\begin{eqnarray} 
 T_{jk} \left (\pjk,\pjk',x_{\pjk'}; M_t-m_{0i}(k_i) \right) &=& 
T_{jk} \left (\pjk,\pjk',x_{\pjk'};\sqrt{m_{0jk}^2(p_{jk}')+\qi^2} \right) 
\\
&& \hskip -5cm +   \int_0^{\infty} d\pjk'' \, \pjk''^2 \int_{-1}^{+1} dx_{\pjk''}  \frac{1}{M_t-m_{0i}(\bqi) - \sqrt{m_{0jk}^2(\pjk'')+\qi^2} }  \,
\cr && \hskip -5cm \times 
T_{jk} \left (\pjk'',\pjk',x_{\pjk'},x_{\pjk''};\sqrt{m_{0jk}^2(\pjk')+\qi^2} \right) T_{\qi} \left (\pjk,\pjk'',x_{\pjk''}; M_t-m_{0i}(\bqi) \right) 
\cr
&& \hskip -5cm -
\Biggl \{
\frac{1}{4} \int_0^{\infty} d\pjk'' \, \pjk''^2 \int_{-1}^{+1} dx_{\pjk''}   \frac{\sqrt{m_{0jk}^2(\pjk')+\qi^2} - \sqrt{m_{0jk}^2(\pjk'')+\qi^2}}{\pjk'^2 - \pjk''^2 } 
\cr && \hskip -5cm \quad \times 
T_{jk} \left (\pjk'',\pjk',x_{\pjk'},x_{\pjk''};\sqrt{m_{0jk}^2(\pjk')+\qi^2} \right) T_{jk} \left (\pjk,\pjk'',x_{\pjk''}; M_t-m_{0i}(\bqi) \right) 
\cr
&& \hskip -5cm
\quad - \frac{1}{2} \int_0^{\infty} d\pjk'' \int_{-1}^{+1} dx_{\pjk''}   \frac{\pjk'^2 \sqrt{m_{0jk}^2(\pjk')+\qi^2}}{\pjk'^2 - \pjk''^2 } 
\cr && \hskip -5cm \quad \times 
T_{jk} \left (\pjk',\pjk',x_{\pjk'},x_{\pjk''};\sqrt{m_{0jk}^2(\pjk')+\qi^2} \right) T_{jk} \left (\pjk,\pjk',x_{\pjk''}; M_t-m_{0i}(k_i) \right) 
\cr
&& \hskip -5cm
\quad - \frac{1}{4} \int_{-1}^{+1} dx_{\pjk''}  \, \pjk' \sqrt{m_{0jk}^2(\pjk')+\qi^2} \, \left (i\pi + \ln \Bigl ( \frac{\pjk''^{max} + \pjk'}{\pjk''^{max} - \pjk'} \Bigr)   \right)
\cr && \hskip -5cm \quad \times 
T_{\qi} \left (\pjk',\pjk',x_{\pjk'},x_{\pjk''};\sqrt{m_{0jk}^2(\pjk')+\qi^2} \right) T_{\qi} \left (\pjk,\pjk',x_{\pjk''}; M_t-m_{0i}(\bqi) \right) 
 \Biggr \}. \nonumber
\label{eq.a8}
\end{eqnarray}
The integral equation is solved for a given value of momentum $\qi$, equal to boost momentum $\k_{jk}$, and for all values of left momentum $\pjk$. 


\begin{acknowledgments}
This work was performed under the auspices of the National Science Foundation
under contract NSF-PHY-1005587 with Ohio University and NSF-PHY-1005501 with the
University of Iowa. Partial support was also provided
by the U. S. Department of Energy, Office of Science of Nuclear Physics, under contract
No. DE-FG02-93ER40756 with Ohio University, and contract
No. DE-FG02-86ER40286 with the University of Iowa. We thank the Ohio
Supercomputer Center (OSC) for the use of
their facilities under grant PHS206. 
\end{acknowledgments}


\bibliography{rel}

\clearpage

%

\begin{table}
\begin{center}
\begin{tabular}{cccccccccccccccccccc}
\hline \hline
 && $E_{nr}$ [MeV] && $E_{r}$ [MeV]&&  $ \frac{E_r-E_{nr}}{E_{nr}}\, [\%] $  \\
\hline 
&& -7.7382 && -7.4825 && +3.3 \\
\hline \hline
&& $E_{r}$ [MeV] && $E_{approx}$ [MeV]&&  $\frac{E_{approx}-E_r}{E_r} \,[
\%] $
\\
\hline
$N=1$ && $-7.4825$ &&  $-7.5412$  &&  $-0.78$
\\
\hline
$C=1$ && $-7.4825$ && $-7.4361$  && $+0.62$
\\
\hline
$N=C=1$ && $-7.4825$ &&  $-7.4934$ && $-0.15$
\\
\hline
$F=1$ && $-7.4825$ &&  $-7.6606$ && $-2.38$
\\
\hline
$G_0^{nr}$ && $-7.4825$ &&  $-7.3446$ && $+1.84$
\\
\hline
$F=1$, $G_0^{nr}$ && $-7.4825$ &&  $-7.4993$ && $-0.22$
\\
\hline \hline
\end{tabular}
\end{center}
\caption{The relativistic (r) and non-relativistic (nr)~\cite{Elster:1998qv} 
three-body binding energies calculated
with the MT-V potential~\cite{Malfliet:1968tj}. 
$E_{approx}$ indicates the relativistic three-body  binding energies 
calculated for Jacobian function $N(\qi,\qi',x_{\qi'})=1$, permutation
coefficient $C(\qi',\qi,x_{\qi'}) =1$,
the analytical term $F(\pjk,\p'jk,\qi)=1$, and using
the non-relativistic free propagator, as well as different combinations thereof.}
\label{table1}
\end{table}

%

\begin{figure}
\begin{center}
\includegraphics[width=15cm]{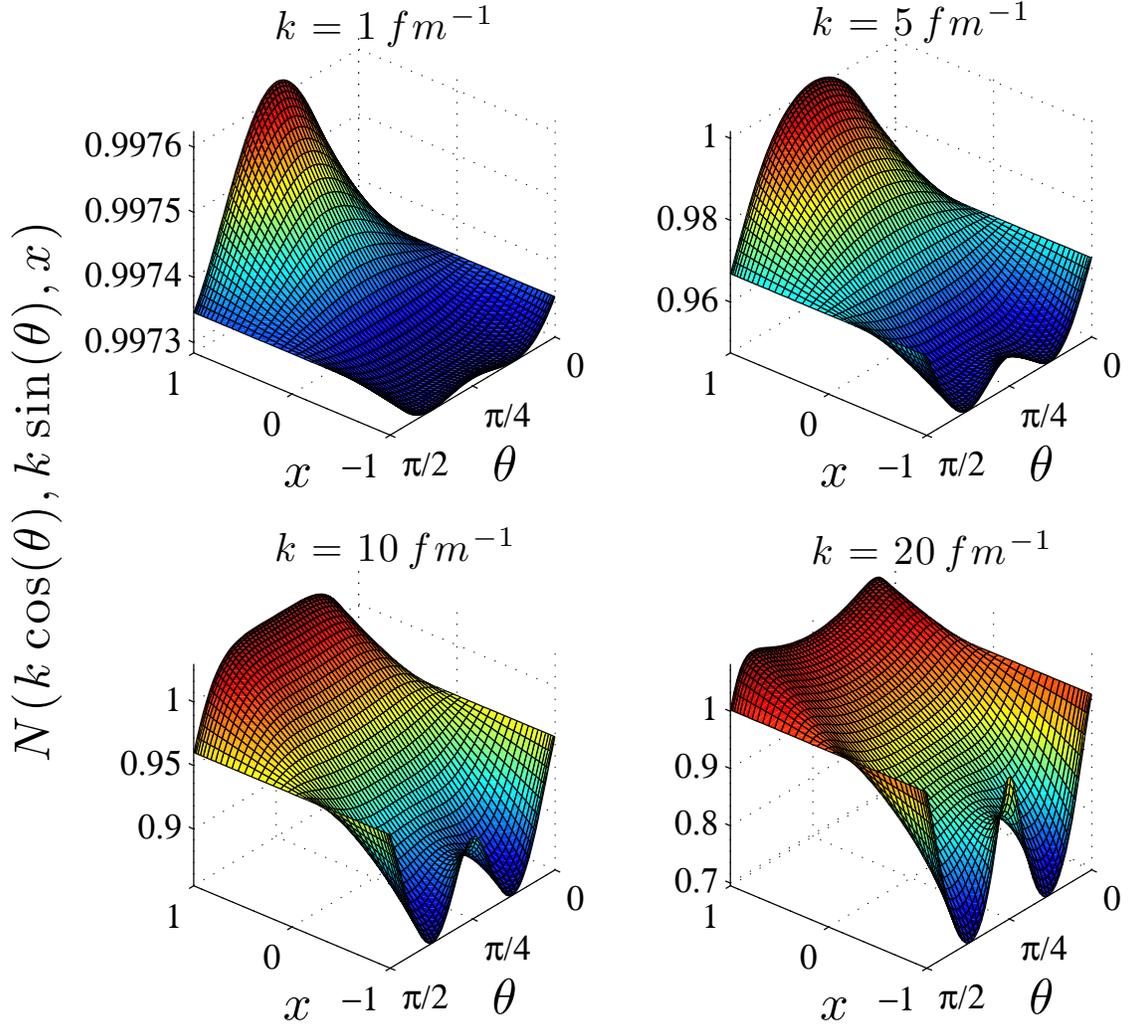}
\end{center}
\caption{(Color online) The matrix elements of Jacobian function
$N(k\cos(\theta),k\sin(\theta),x)$, Eq.~(\ref{eq.4.25}), as function of the
angles x and $\theta$ calculated for different values of momentum $k$.}
\label{fig2}
\end{figure}

\begin{figure}
\begin{center}
\includegraphics[width=15cm]{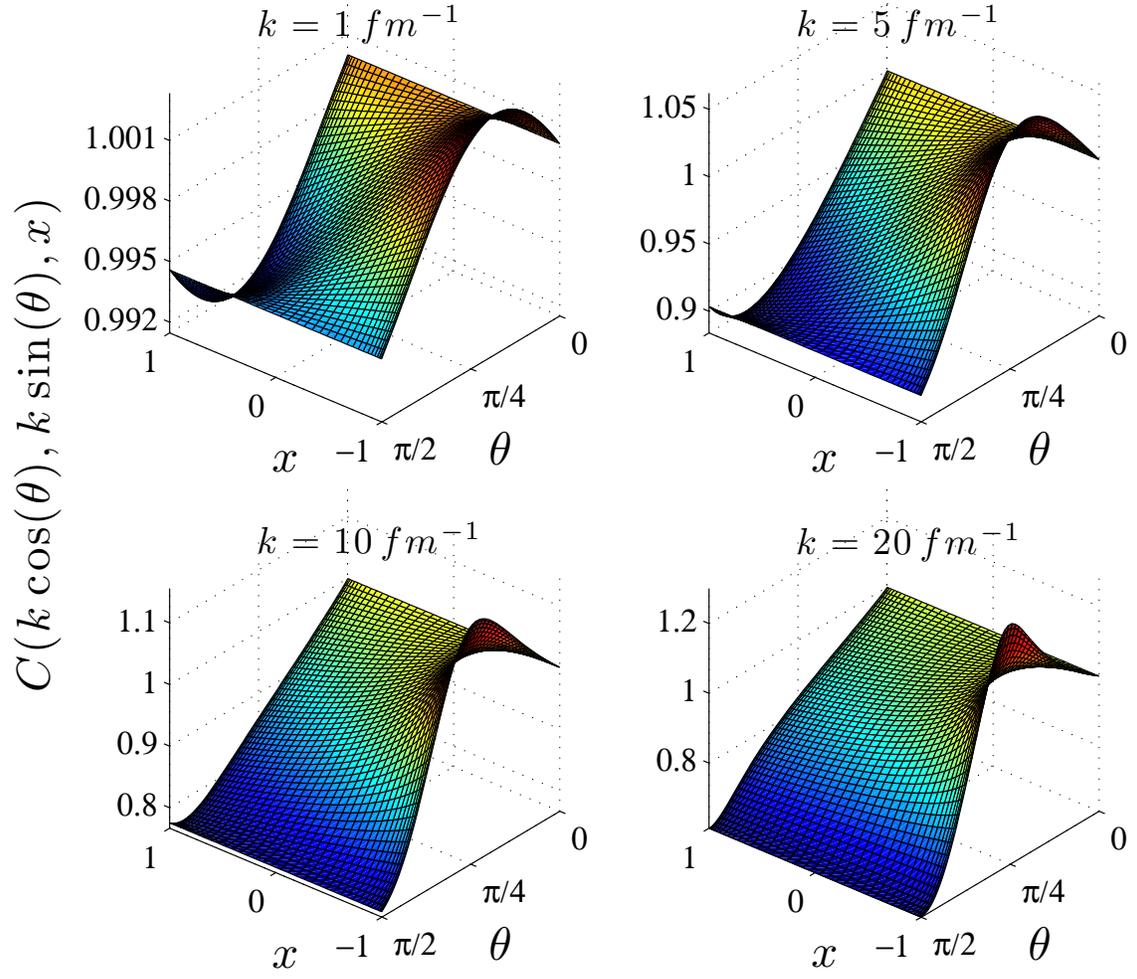} 
\end{center}
\caption{(Color online) The matrix elements of permutation coefficient
$C(k\cos(\theta),k\sin(\theta),x)$, Eq.~(\ref{eq.4.29}), as function of the
angles x and $\theta$ calculated for different values of momentum $k$.}
\label{fig3}
\end{figure}

\begin{figure}
\begin{center}
\includegraphics[width=15cm]{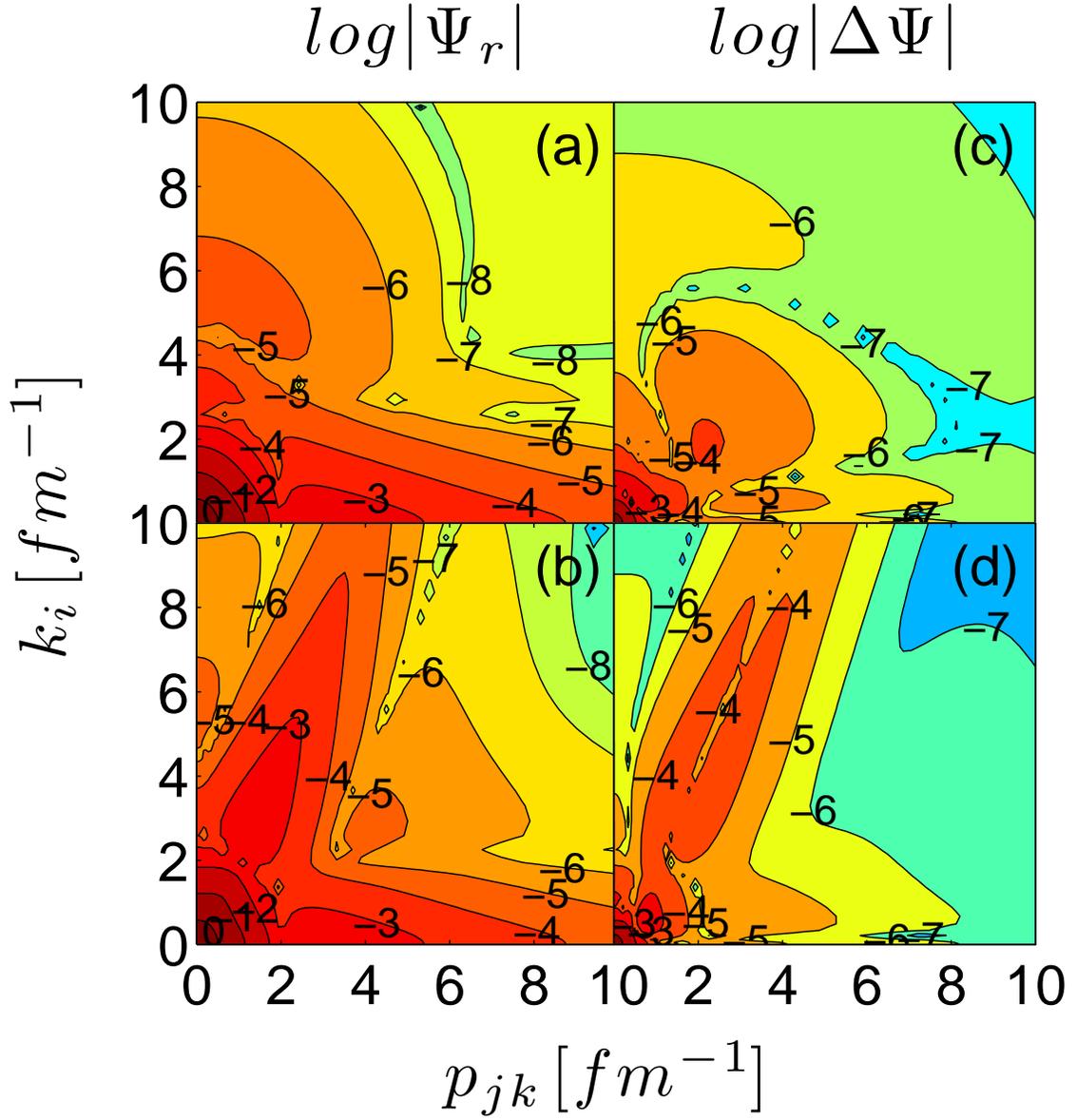}
\end{center}
\caption{(Color online) The  magnitude of the relativistic three-body 
 bound state wave function $\Psi(\pjk,\qi,x_{\pjk})$ as function of the
pair momentum $\pjk$ and the spectator momentum $\qi$ for fixed values
 $x_{\pjk}=0$ (a) and $x_{\pjk}=+1$ (b) obtained with
the MT-V potential~\cite{Malfliet:1968tj,Elster:1998qv}. 
The difference between relativistic and non-relativistic wave
functions are  shown for fixed values
 $x_{\pjk}=0$ in panel (c) and $x_{\pjk}=+1$ in (d).}
\label{fig4}
\end{figure}

\begin{figure}
\begin{center}
\includegraphics[width=14cm]{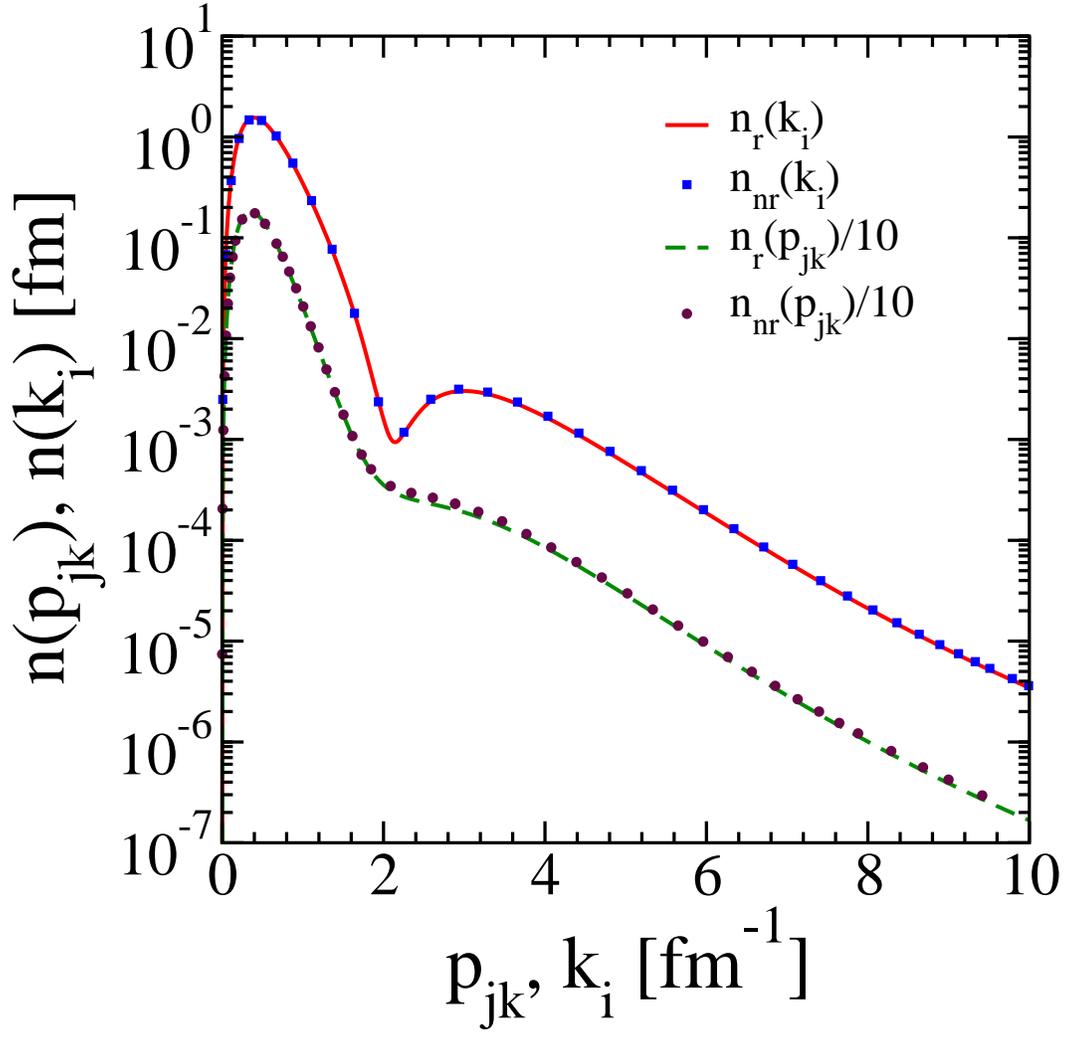}
\end{center}
\caption{(Color online) The relativistic and non-relativistic momentum
distribution function $n(\qi)$(solid and filled squares) and $n(\pjk)$
(dashed and filled circles) 
 obtained  from the MT-V potential~\cite{Malfliet:1968tj,Elster:1998qv}. 
The $n(\pjk)$ are scaled with a factor 0.1.}
\label{fig5}
\end{figure}

\begin{figure}
\begin{center}
\includegraphics[width=12cm]{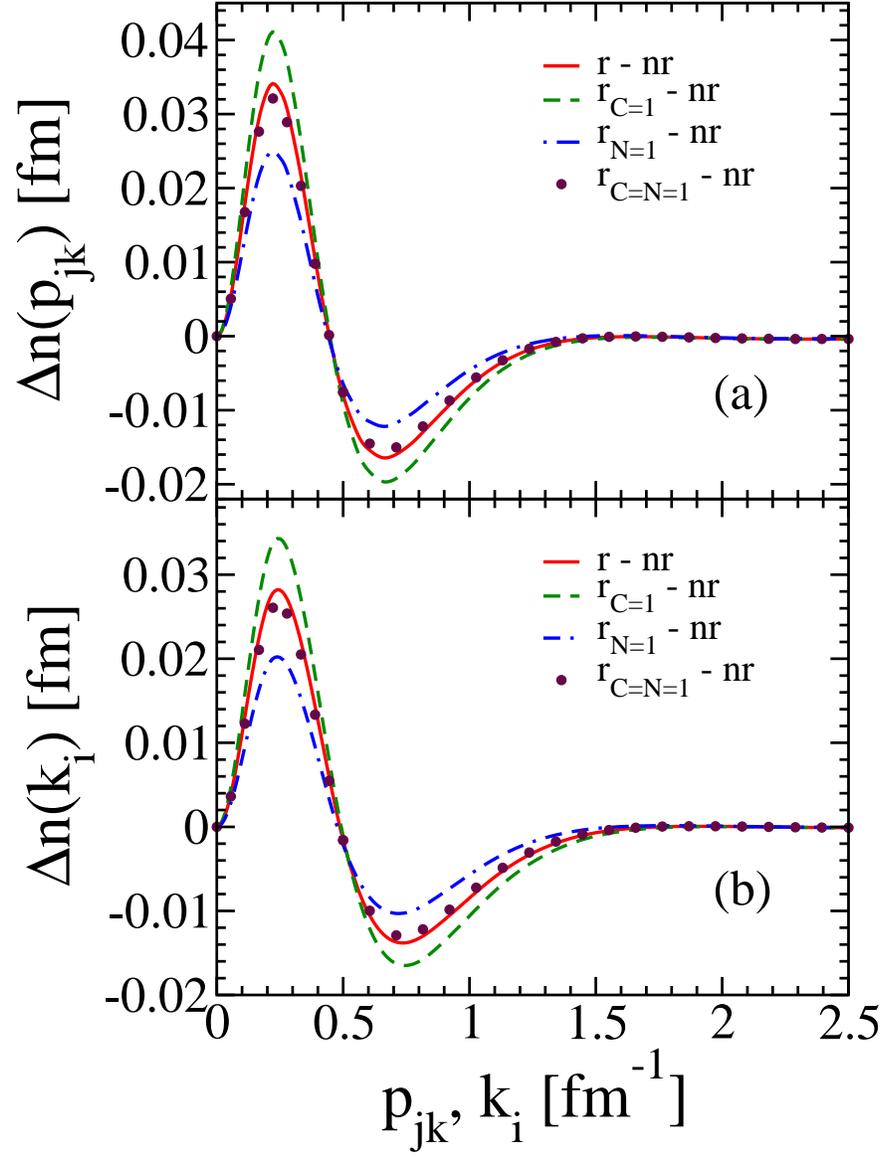}
\end{center}
\caption{(Color online) The difference between the relativistic and 
non-relativistic momentum distribution functions
(a) for $n(\pjk)$ and (b) for $n(\qi)$ calculated from the MT-V potential
(solid line). The dashed line shows the difference for the case when $C=1$
in the relativistic calculation, while for the dash-dotted line $N=1$. The
filled circles represent a calculation in in which both, $C=N=1$.
}
\label{fig6}
\end{figure}

\end{document}